%% file: cellfree.tex
\documentclass{article}
\usepackage{graphicx}
\usepackage{amsmath}
\usepackage{cite}
\usepackage{algorithm}
\usepackage{algpseudocode}

\begin{document}

\title{Near-Optimal Pilot Assignment in Cell-Free Massive MIMO}

\author{Raphael~M.~Guedes$^1$\\
José~F.~de~Rezende$^2$\\
Valmir~C.~Barbosa$^{2,3,}$\thanks{Corresponding author
(valmircbarbosa@gmail.com.)}\\
\\
$^1$Informatics and Computer Science Department\\
State University of Rio de Janeiro\\
Rio de Janeiro, RJ 20550-900, Brazil\\
$^2$Systems Engineering and Computer Science Program\\
Federal University of Rio de Janeiro\\
Rio de Janeiro, RJ 21941-914, Brazil\\
$^3$Graduate Program in Computational Sciences\\
State University of Rio de Janeiro\\
Rio de Janeiro, RJ 20550-900, Brazil}

\date{}

\maketitle

\begin{abstract}
Cell-free massive MIMO systems are currently being considered as potential
enablers of future (6G) technologies for wireless communications. By combining
distributed processing and massive MIMO, they are expected to deliver improved
user coverage and efficiency. A possible source of performance degradation in
such systems is pilot contamination, which contributes to causing interference
during uplink training and affects channel estimation negatively. Contamination
occurs when the same pilot sequence is assigned to more than one user. This is
in general inevitable, as the number of mutually orthogonal pilot sequences
corresponds to only a fraction of the coherence interval. We introduce a new
algorithm for pilot assignment and analyze its performance both from a
theoretical perspective and in computational experiments. We show that it has an
approximation ratio close to 1 for a plausibly large number of orthogonal pilot
sequences, as well as low computational complexity under massive parallelism. We
also show that, on average, it outperforms other methods in terms of per-user
SINR and throughput on the uplink.

\end{abstract}

\newpage
\section{Introduction}
\label{intr}

We consider a cell-free massive MIMO system as described in \cite{naylm17},
which is characterized by a large number $M$ of single-antenna, geographically
distributed APs simultaneously serving $K\ll M$ autonomous users via a TDD
scheme. Each coherence interval, assumed to be of duration $\tau_\mathrm{c}$
(samples), is divided into a phase for uplink training and two others for
downlink and uplink data transmission. Training refers to the sending by each
user to all APs of a $\tau_\mathrm{p}$-sample pilot sequence (a pilot), with
$\tau_\mathrm{p}\ll\tau_\mathrm{c}$, used by each AP to estimate the channel for
subsequent downlink and uplink data transmission for that user. The APs are
capable of computationally efficient signal processing, and are moreover
connected to a CPU by a fronthaul network. Two tasks the CPU handles are pilot
assignment and power allocation.

Our goal is to contribute to the development of algorithms for pilot assignment.
Before we continue, however, it is important to note that the development of
cell-free massive MIMO has continued to evolve since the publication of
\cite{naylm17}, aiming to both enlarge the physical capabilities of the system
(e.g., by providing each AP with multiple antennas and expanding the system's
computational capacity) and to more realistically address some performance
bottlenecks and other difficulties that were not contemplated at the time. These
have included synchronization issues related to TDD \cite{gsl21}, reciprocity
calibration to make possible the intended use of the same channel for both
downlink and uplink traffic \cite{kyhnc22}, and scalability \cite{bs20}.

In this letter, we assume that all available pilots are orthogonal to one
another. Thus, given the number of samples $\tau_\mathrm{p}$ in a pilot,
the number of pilots is $P=\tau_\mathrm{p}$. Assigning pilots to users can be
complicated if $P<K$, since in this case at least two users must be assigned the
same pilot. This gives rise to so-called pilot contamination, whose consequence
is a reduced data rate for the users involved. In \cite{naylm17}, the channel
between AP $m$ and user $k$ is modeled as
\begin{equation}
g_{mk}=\beta_{mk}^{1/2}h_{mk},
\end{equation}
where $\beta_{mk}$ is the large-scale fading and $h_{mk}$ is the small-scale
fading. The $\beta_{mk}$'s are assumed to remain constant during each coherence
interval and the $h_{mk}$'s to be i.i.d.\ $\mathcal{CN}(0,1)$ random variables.
Estimating channel $g_{mk}$ during uplink training causes a pilot-contamination
effect on $k$ proportional to $\sum_{k'\in U_k\setminus\{k\}}g_{mk'}$, where
$U_k$ is the set of users assigned the same pilot as user $k$ (itself included).
The variance of this quantity relative to the $h_{mk}$'s, after totaled over all
APs, is given by
\begin{equation}
v_k=\sum_{k'\in U_k\setminus\{k\}}\sum_{m=1}^M\beta_{mk'}.
\label{var}
\end{equation}

Variance $v_k$ is therefore fundamentally tied to the issue of pilot
contamination, so minimizing it during pilot assignment plays a central role in
attenuating the deleterious effects of pilot scarcity on $k$. Globally, the
problem to be solved can be formulated as finding a partition of the set of
users into $P$ subsets, aiming to assign the same pilot to all users in the same
subset. The goal is to find a partition $\mathcal{P}=\{S_1,\ldots,S_P\}$ that
minimizes $\sum_{S\in\mathcal{P}}\sum_{k\in S}v_k$, where
\begin{align}
\sum_{k\in S}v_k
&= \sum_{k\in S}\sum_{k'\in S\setminus\{k\}}\sum_{m=1}^M\beta_{mk'} \\
&= (\vert S\vert-1)\sum_{k\in S}\sum_{m=1}^M\beta_{mk}.
\label{sumSvk}
\end{align}
This is an NP-hard optimization problem, but here we demonstrate that it can be
tackled by a greedy algorithm so that the optimum is approximated to within a
ratio that improves as the number of pilots $P$ increases.

We proceed as follows. In Section~\ref{contr}, we briefly review the relevant
state of the art and relate our contribution to it. We then recap a few system
model details in Section~\ref{smod}, where we continue to follow \cite{naylm17}
closely. In Section~\ref{alg}, we recast the problem of finding $\mathcal{P}$ to
minimize $\sum_{S\in\mathcal{P}}\sum_{k\in S}v_k$ in graph-theoretic terms, and
describe and analyze our near-optimal algorithm to solve it. Computational
results are given in Section~\ref{rslt} and we conclude in Section~\ref{concl}.

\section{State of the art and contribution}
\label{contr}

Two baseline approaches to pilot assignment are RANDOM, which assigns a pilot
chosen uniformly at random to each user, and GREEDY \cite{naylm17}, which begins
as RANDOM and then repeatedly identifies the user $k$ for which a performance
measure of choice is worst and replaces its pilot so that variance $v_k$ is
minimized. The latter goes on while the selected user's pilot does indeed
change.  More elaborate approaches from recent years include Improved BASIC
(IBASIC) \cite{qzj22} and some that use graph theory-based techniques
\cite{lzja20,bdfzf21,zhlw21}. IBASIC first sorts the users in descending order
of $\sum_{m=1}^M\beta_{mk}$ and assigns pilots to the first $P$ users as in
RANDOM. It then goes through the succeeding $K-P$ users, in order, each of which
gets assigned the pilot that currently gets closest to minimizing
$\sum_{k'\in U_k\setminus\{k\}}\beta_{mk'}$ while respecting a preestablished
maximum number of users $\delta$ per pilot. AP $m$ is the one for which
$\beta_{mk}$ is greatest.

The approaches in \cite{lzja20,bdfzf21,zhlw21} aim to pose the problem of pilot
assignment in terms of an undirected graph whose vertices are the $K$ users. All
three aim to obtain a $P$-set partition of the vertex set, but the ones in
\cite{lzja20,bdfzf21}, based respectively on vertex coloring (COLORING) and on
finding a maximum-weight matching on a bipartite graph (MATCHING), take
circuitous routes to their goals and seem oblivious to the precise definition of
partition $\mathcal{P}$ given in Section~\ref{intr}. COLORING operates by
repeatedly adjusting the graph's density, based on the $\beta_{mk}$'s, until it
becomes $P$-colorable according to a heuristic. MATCHING, in turn, iterates
until either a performance criterion is met or a preestablished maximum number
of iterations is reached. In each iteration, the $\beta_{mk}$'s are used to
create a bipartite graph in which $P$ of the users are assigned pilots and the
remaining $K-P$ are to share pilots with them based on the resulting
maximum-weight matching.

In our view, COLORING and MATCHING are both based on a failure to realize that
the most direct route to finding partition $\mathcal{P}$ is to also consider the
perspective that is dual to the minimization involved in the partition's
definition. Such dual perspective is that of maximization: to find partition
$\mathcal{P}$, look for a maximum-weight $P$-cut of an edge-weighted complete
graph on $K$ vertices whose weights depend on the $\beta_{mk}$'s. A $P$-cut is
simply the set of all edges connecting vertices from different sets of
$\mathcal{P}$. The approach in \cite{zhlw21}, known as Weighted Graphic
Framework (WGF), is on the other hand firmly in line with this idea but uses
edge weights that are essentially unjustified. WGF uses the maximum-weight
$P$-cut algorithm from \cite{sg76} directly: it starts by initializing the $P$
sets of the partition by adding an arbitrarily selected user to each of them; it
then considers each of the remaining $K-P$ users (say $k$), computes for each
set the total internal edge weight it will have if $k$ is added to it, and
finally adds $k$ to the set whose weight will be minimum. The total weight of
the $P$-cut output by this algorithm accounts for a fraction of the optimal
total weight of at least $(P-1)/P$ \cite{sg76}, so WGF has an approximation
ratio that approaches $1$ as $P$ increases.

In this letter, we pick up where WGF left off and contribute a new algorithm to
assign pilots to users. Like WGF, this algorithm looks for a maximum-weight
$P$-cut on an edge-weighted complete graph on the $K$ users. Unlike WGF, though,
edge weights stay true to the principle of reflecting the variance of the
interference caused by pilot contamination during uplink training, as discussed
in Section~\ref{intr}. We call the new algorithm Greedy Edge Contraction (GEC)
and prove that it too has an approximation-ratio lower bound that approaches $1$
as $P$ increases, now given by $(P-1)/(P+1)$. In this sense, both WGF and GEC
are near-optimal, with WGF more so, though only slightly for relatively large
$P$, since $(P-1)/P>(P-1)/(P+1)$. This difference notwithstanding, our results
in Section~\ref{rslt} show that GEC performs better than other methods,
including WGF. As we will see, this is only partly due to the poorly defined
edge weights that WGF uses.

\section{System model essentials}
\label{smod}

\begin{figure}[t]
\centering
\includegraphics[scale=0.65]{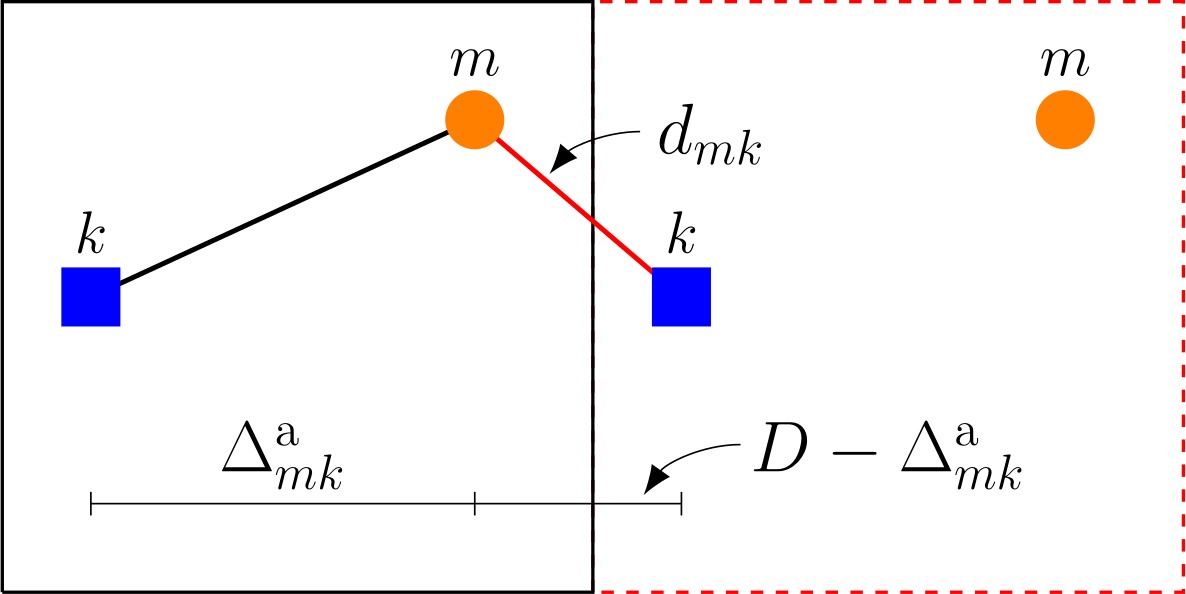}
\caption{Attaching a ``phantom'' copy of the original $D\times D$ region to its
right-hand boundary allows the two choices of displacement along the abscissas
to be visualized. In this example, clearly
$D-\Delta_{mk}^\mathrm{a}<\Delta_{mk}^\mathrm{a}$, so $D-\Delta_{mk}^\mathrm{a}$
should be preferred, yielding $d_{mk}$ as indicated.}
\label{figwrap}
\end{figure}

We assume APs and users to be placed in a $D\times D$ square region, at
coordinates $(x_i,y_i)$ for $i$ an AP or a user. We also assume that this region
wraps itself around the boundaries on both dimensions. For AP $m$ and user $k$,
letting $\Delta_{mk}^\mathrm{a}=\vert x_m-x_k\vert$ and likewise
$\Delta_{mk}^\mathrm{o}=\vert y_m-y_k\vert$ implies that the distance $d_{mk}$
between them is such that
\begin{equation}
d_{mk}^2=
{\textstyle\min^2}\{\Delta_{mk}^\mathrm{a},D-\Delta_{mk}^\mathrm{a}\}+
{\textstyle\min^2}\{\Delta_{mk}^\mathrm{o},D-\Delta_{mk}^\mathrm{o}\}.
\end{equation}
This has become customary in the field (see, e.g.,
\cite{naylm17,bs20,lzja20,bdfzf21,qzj22}) and aims to help attenuate the
inevitable boundary effects that come with a finite connected region. The idea
is to generalize the Euclidean distance formula on the plane,
$d_{mk}^2=(\Delta_{mk}^\mathrm{a})^2+(\Delta_{mk}^\mathrm{o})^2$, by allowing
each of the two squared displacements on the right-hand side to be replaced by
its over-the-boundary version, $(D-\Delta_{mk}^\mathrm{a})^2$ or
$(D-\Delta_{mk}^\mathrm{o})^2$ as the case may be, whenever that leads to a
smaller $d_{mk}$. A simple example where wrapping is used only along the
abscissas is given in Figure~\ref{figwrap}.

For $d_0,d_1$ (m) the reference distances, $f$ (MHz) the carrier frequency, and
$h_\mathrm{AP}$, $h_\mathrm{user}$ (m) the antenna heights, the path loss
$\mathrm{PL}_{mk}$ (dB) corresponding to $d_{mk}$ follows the same three-slope
model as \cite{naylm17}, given by
\begin{equation}
\mathrm{PL}_{mk}=\begin{cases}
-L-15\ell_1-20\ell_0,
& \text{if }d_{mk}\le d_0, \\
-L-15\ell_1-20\ell_{mk},
& \text{if }d_0<d_{mk}\le d_1, \\
-L-35\ell_{mk},
& \text{if }d_1<d_{mk},
\end{cases}
\end{equation}
where
\begin{equation}
\begin{split}
L
&= 46.3+33.9\log_{10}f-13.82\log_{10}h_\mathrm{AP} \\
&\quad -(1.1\log_{10}f-0.7)h_\mathrm{user}+1.56\log_{10}f-0.8,
\end{split}
\end{equation}
$\ell_{mk}=\log_{10}d_{mk}$, $\ell_0=\log_{10}d_0$, and $\ell_1=\log_{10}d_1$.
The resulting large-scale fading is
\begin{equation}
\beta_{mk}=10^{10^{-1}(\mathrm{PL}_{mk}+\sigma_\mathrm{sf}z_{mk})},
\end{equation}
where $\sigma_\mathrm{sf}$ (dB) is the shadow-fading standard deviation and
$z_{mk}$ is an $\mathcal{N}(0,1)$ random variable. We assume that the $z_{mk}$'s
are uncorrelated with one another and that the $\beta_{mk}$'s are available
whenever needed.

As in \cite{naylm17}, we assume that each AP calculates MMSE estimates of the
channels between itself and the users from a combination of all users' pilots
sent to it during training. The estimated channel between AP $m$ and user $k$
has expected gain
\begin{equation}
\gamma_{mk}=
\frac
{\tau_\mathrm{p}\rho_\mathrm{p}\beta_{mk}^2}
{\tau_\mathrm{p}\rho_\mathrm{p}\sum_{k'\in U_k\setminus\{k\}}\beta_{mk'}+1}.
\end{equation}
Using the notations
\begin{align}
a_{kk'}
&= \left(\sum_{m=1}^M \gamma_{mk}\frac{\beta_{mk'}}{\beta_{mk}}\right)^2, \\
b_{kk'}
&= \sum_{m=1}^M\gamma_{mk}\beta_{mk'}, \\
c_k
&= \rho_\mathrm{u}^{-1}\sum_{m=1}^M\gamma_{mk},
\end{align}
the resulting SINR on the uplink is given by
\begin{equation}
\mathrm{SINR}_k^\mathrm{u}=
\frac
{\eta_k\left(\sum_{m=1}^M\gamma_{mk}\right)^2}
{\sum_{k'\in U_k\setminus\{k\}}
\eta_{k'}a_{kk'}+\sum_{k'=1}^K\eta_{k'}b_{kk'}+c_k}.
\label{sinr}
\end{equation}
In the expressions for $\gamma_{mk}$ and $c_k$, $\rho_\mathrm{p}$ and
$\rho_\mathrm{u}$ are the normalized uplink SNR for training and for data
transmission, respectively. The resulting throughput for user $k$ is
\begin{equation}
R_k^\mathrm{u}=
\frac{B}{2}
\left(1-\frac{\tau_\mathrm{p}}{\tau_\mathrm{c}}\right)
\log_2(1+\mathrm{SINR}_k^\mathrm{u}),
\label{rate}
\end{equation}
where $B$ (Hz) is the channel bandwidth. In this equation, the factor
$(1-\tau_\mathrm{p}/\tau_\mathrm{c})/2$ serves first to deduct the fraction of
the coherence interval that is used for pilot transmission, then to further
deduct half of what remains, which we assume is reserved for data transmission
on the downlink.

Eq.~(\ref{sinr}) is central to the comparative computational study we carry out
in Section~\ref{rslt}, so the $\eta_k$'s appearing in it, which work as power
control coefficients, must be determined for each new assignment of pilots to
users. As customary, in order to ensure fairness toward all users we express
power allocation as the max-min problem, on variables $t$ and
$\eta_1,\ldots,\eta_K$, given by
\begin{align}
\text{maximize } &t \\
\text{subject to }
&t\le\mathrm{SINR}_k^\mathrm{u}, &k=1,\ldots,K,
\label{cvx1} \\
&0\le\eta_k\le 1, &k=1,\ldots,K.
\label{cvx2}
\end{align}
This is a quasilinear problem, so we do bisection on variable $t$ to solve it,
tackling only the linear feasibility program given by Eqs.~(\ref{cvx1})
and~(\ref{cvx2}) for each fixed value of $t$. The resulting
$\mathrm{SINR}_k^\mathrm{u}$ is necessarily the same for every user $k$. Thus,
whenever referring to these SINR values or the corresponding throughputs, we
henceforth use simply $\mathrm{SINR^u}$ and $R^\mathrm{u}$, respectively.

\section{The GEC algorithm}
\label{alg}

Like WGF, GEC does pilot assignment to users by solving the MAX $P$-CUT problem
on an edge-weighted complete graph, now denoted by $G_K$, having a vertex set
that corresponds to the set of users. MAX $P$-CUT asks that the vertex set of
$G_K$ be partitioned into $P$ sets in such a way that the sum of the weights of
all inter-set edges is maximized, or equivalently the sum over all intra-set
edges is minimized.

The idea is for each of these $P$ sets to correspond to a set of users to which
the same pilot is assigned. It is therefore crucial that weights be selected in
a way that relates directly and clearly to the potential for pilot contamination
between the users in question. In line with our reasoning in Section~\ref{intr},
we quantify some user $k$'s contribution to the pilot-contamination effect on
each of the users it shares the pilot with as $\beta_k$, henceforth defined as
\begin{equation}
\beta_k=\sum_{m=1}^M\beta_{mk}.
\end{equation}
Thus, the weight of the edge interconnecting vertices $i$ and $j$ in $G_K$,
denoted by $w_{ij}$, is
\begin{equation}
w_{ij}=\beta_k+\beta_{k'},
\label{weight-init}
\end{equation}
assuming that vertex $i$ corresponds to user $k$ and vertex $j$ to user $k'$ (or
$i$ to $k'$, $j$ to $k$).

MAX 2-CUT is one of the classic NP-hard problems, so the trivially more general
MAX $P$-CUT is NP-hard as well. We approach its solution by employing the
generalization given in \cite{kkbh08} of their own MAX 2-CUT algorithm. The
resulting GEC runs for $K-P$ iterations, each consisting in the contraction of
an edge, say $(i^*,j^*)$, thus joining vertices $i^*$ and $j^*$ into a single
new vertex, say $\ell$, and moreover connecting to $\ell$ every vertex
previously connected to $i^*$ or $j^*$.

These iterations result in a sequence of graphs that, like the initial $G_K$,
are also edge-weighted complete graphs. Unlike $G_K$, however, vertices in these
graphs are no longer necessarily identified with single users, but generally
with non-singleton sets of users as well. The last graph in the sequence,
denoted by $G_P$, has $P$ vertices, one for each pilot.

The general formula for the weight $w_{ij}$ between vertices $i$ and $j$, valid
for all graphs in the sequence, is
\begin{align}
w_{ij}
&= \sum_{k\in S_i}\sum_{k'\in S_j}\beta_k
+\sum_{k\in S_j}\sum_{k'\in S_i}\beta_k \\
&= n_j\sum_{k\in S_i}\beta_k+n_i\sum_{k\in S_j}\beta_k,
\label{weight-general}
\end{align}
where $S_i$ is the set of users to which vertex $i$ corresponds and $n_i$ is its
size. This expression generalizes the one in Eq.~(\ref{weight-init}), which
refers to an edge in $G_K$ with $S_i=\{k\}$ and $S_j=\{k'\}$ (or vice versa). In
order for the formula in Eq.~(\ref{weight-general}) to remain valid as vertices
$i^*$ and $j^*$ are joined to form vertex $\ell$, it suffices that each edge
$(i,\ell)$ such that $i\neq i^*,j^*$ be given weight
$w_{i\ell}=w_{ii^*}+w_{ij^*}$, that is, the sum of the weights of the two edges
that used to connect $i$ to $i^*$ and $j^*$ before the contraction of edge
$(i^*,j^*)$. Note also that summing up the weights of all $S_i$'s intra-set
edges yields
\begin{equation}
\sum_{k\in S_i}\sum_{\genfrac{}{}{0pt}{}{k'\in S_i}{k'\neq k}}\beta_k=
(n_i-1)\sum_{k\in S_i}\beta_k,
\label{contamination}
\end{equation}
which as expected is simply a rewrite of Eq.~(\ref{sumSvk}). The sum of this
quantity over all vertices (every $i$) is what is targeted for minimization as
the solution to MAX $P$-CUT is approximated by GEC. The heart of GEC at each
iteration is therefore to select for contraction the edge of least weight. GEC
is summarized as the pseudocode in Algorithm~\ref{alggec}.

\begin{algorithm}[t]
\caption{Pseudocode for GEC.}
\label{alggec}
\textbf{Input:} $G_K$, edge weights $w_{ij}$ as in Eq.~(\ref{weight-init}) \\
\textbf{Output:} $G_P$
\begin{algorithmic}[1]
\State $G\leftarrow G_K$
\State $n\leftarrow K$
\While {$n>P$} \label{while}
\State Let $(i^*,j^*)$ be a minimum-weight edge of $G$ \label{costliest}
\State $S\leftarrow S_{i^*}\cup S_{j^*}$
\For {$i\neq i^*,j^*$} \label{for1}
\State $w^{(i)}\leftarrow w_{ii^*}+w_{ij^*}$
\EndFor
\State Contract edge $(i^*,j^*)$ by joining vertices $i^*$ and $j^*$ into a new
vertex $\ell$ \label{ctr}
\State $S_\ell\leftarrow S$
\For {$i\neq\ell$} \label{for2}
\State $w_{i\ell}\leftarrow w^{(i)}$
\EndFor
\State $n\leftarrow n-1$
\EndWhile
\State $G_P\leftarrow G$
\end{algorithmic}
\end{algorithm}

An extension of the analysis in \cite{kkbh08} reveals that
\begin{equation}
W^\mathrm{obt}\ge\frac{P-1}{P+1}\,W^\mathrm{opt},
\end{equation}
where $W^\mathrm{obt}$ is the total weight of the edges of $G_P$ (i.e., the
total weight of the obtained $P$-cut of $G_K$) and $W^\mathrm{opt}$ is its
optimal value. To see that this holds, let $W_K$ be the total weight of the
edges of $G_K$ and then use Lemma~1 from \cite{kkbh08}, which is valid for
MAX $P$-CUT as much as it is for MAX 2-CUT. It states that
\begin{equation}
W^\mathrm{ctr}\le\frac{2(K-P)}{(K-1)(P+1)}\,W_K,
\label{lemma1}
\end{equation}
where $W^\mathrm{ctr}$ is the total weight of the $K-P$ edges contracted during
the iterations. Using Eq.~(\ref{lemma1}) and the fact that
$W_K\ge W^\mathrm{opt}$, we obtain
\begin{align}
W^\mathrm{obt} &= W_K-W^\mathrm{ctr} \\
&\ge W_K-\frac{2(K-P)}{(K-1)(P+1)}\,W_K \\
&\ge \frac{(K-1)(P+1)-2(K-P+P-1)}{(K-1)(P+1)}\,W_K \\
&\ge \frac{(K-1)(P-1)}{(K-1)(P+1)}\,W^\mathrm{opt} \\
&= \frac{P-1}{P+1}\,W^\mathrm{opt}. 
\end{align}

This means that GEC, similarly to WGF (see Section~\ref{contr}), is capable of
approximating the optimal $P$-cut of $G_K$ so long as the number $P$ of pilots
is sufficiently large. For example, with $P=25$ we get
$W^\mathrm{obt}\ge 0.92\,W^\mathrm{opt}$ for GEC and
$W^\mathrm{obt}\ge 0.96\,W^\mathrm{opt}$ for WGF. This might seem to put WGF at
an advantage over GEC, perhaps one counterbalanced by GEC's edge weights being
well-founded while those of WGF are not. What we have observed is more nuanced
than this, though, as we discuss in Section~\ref{concl}.

As for GEC's computational complexity, note that its costliest step is the one
in line~\ref{costliest}, which requires $O(K^2)$ time, followed by the loop in
line~\ref{for1}, line~\ref{ctr}, and the loop in line~\ref{for2}, each running
in $O(K)$ time. Considering that the loop in line~\ref{while} repeats $O(K)$
times, the overall time required by GEC on a sequential device is $O(K^3)$.
However, so long as ASICs can be designed to provide the necessary massive
parallelism, the time requirement of line~\ref{costliest} can be lowered to
$O(\log K)$ (see, e.g., \cite{absc15} and references therein). Likewise, the
loop in line~\ref{for1}, as well as line~\ref{ctr} and the loop in
line~\ref{for2}, can much more easily be sped up to run in $O(1)$ time. The
overall time required by GEC can therefore be reduced to $O(K\log K)$. This
remains unaltered if we add the time for calculating the $\beta_k$'s, whenever
the $\beta_{mk}$'s change, prior to running GEC. Once again assuming the
necessary massive parallelism, this can be achieved in $O(\log M)$ time, which
gets reduced to $O(\log K)$ for $M=aK$ with $a$ a constant. Since by assumption
we have $K\ll M$, for consistency we require only that $a>1$ (we use $a=4$ for
our computational results). WGF runs faster on a sequential computer, requiring
$O(K^2)$ time, but assuming massive parallelism reduces this to the same
$O(K\log K)$ time as in the case of GEC. This is owed to the fact that both the
factor $O(K^2)$ for GEC to obtain a minimum, and $O(K)$ for WGF, get reduced to
$O(\log K)$.

\section{Computational results}
\label{rslt}

We use the parameter values given in Table~\ref{tab1}, where the value of
$\rho_\mathrm{p},\rho_\mathrm{u}$ is for the channel bandwidth $B$ in the table,
a transmit power of $0.1$~W, a temperature of $290$~K, and a noise figure of
$9$~dB. Each value of $\tau_\mathrm{c}$ is compatible either with mobile users
at highway speeds ($\tau_\mathrm{c}=750$; see Table~2.1 in \cite{mlyn16}) or
with users at urban-road speeds ($\tau_\mathrm{c}=1000,1250$, extending that
same table for a speed of at most $18$ m/s). We use $M=400$ and $K=100$
throughout.

\begin{table}[t] 
\caption{System model parameters.}
\label{tab1}
\centering
\input{table1}
\end{table}

\begin{figure}[t]
\centering
\includegraphics[scale=0.65]{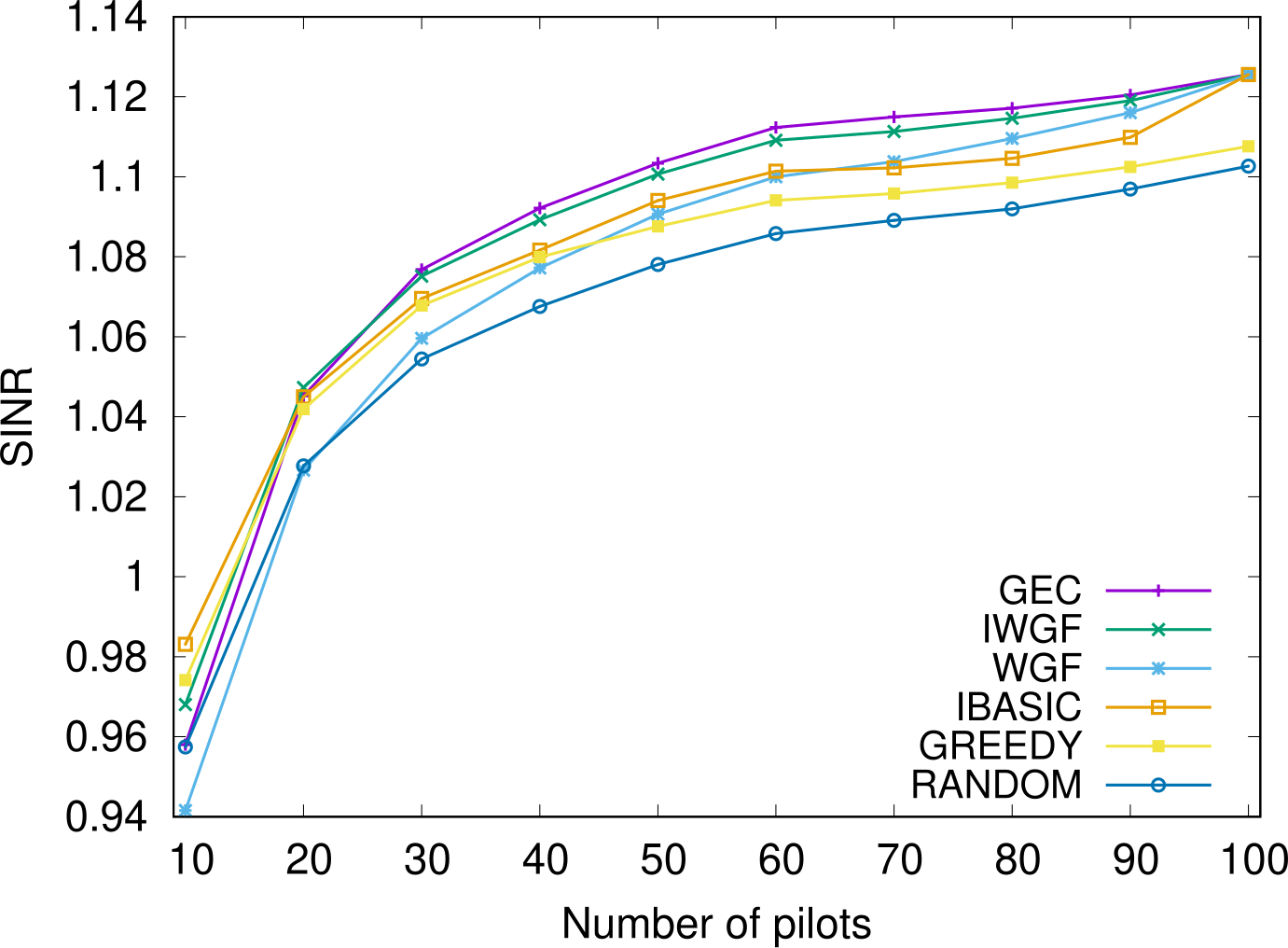}
\caption{$\mathrm{SINR}^\mathrm{u}$ vs.\ number of pilots $P$.
Confidence-interval bounds at the $95\%$ level are about $\pm 0.4\%$ of the
average and occur for WGF at $P=10$.}
\label{figsinr}
\end{figure}

\begin{figure}[p]
\centering
\includegraphics[scale=0.65]{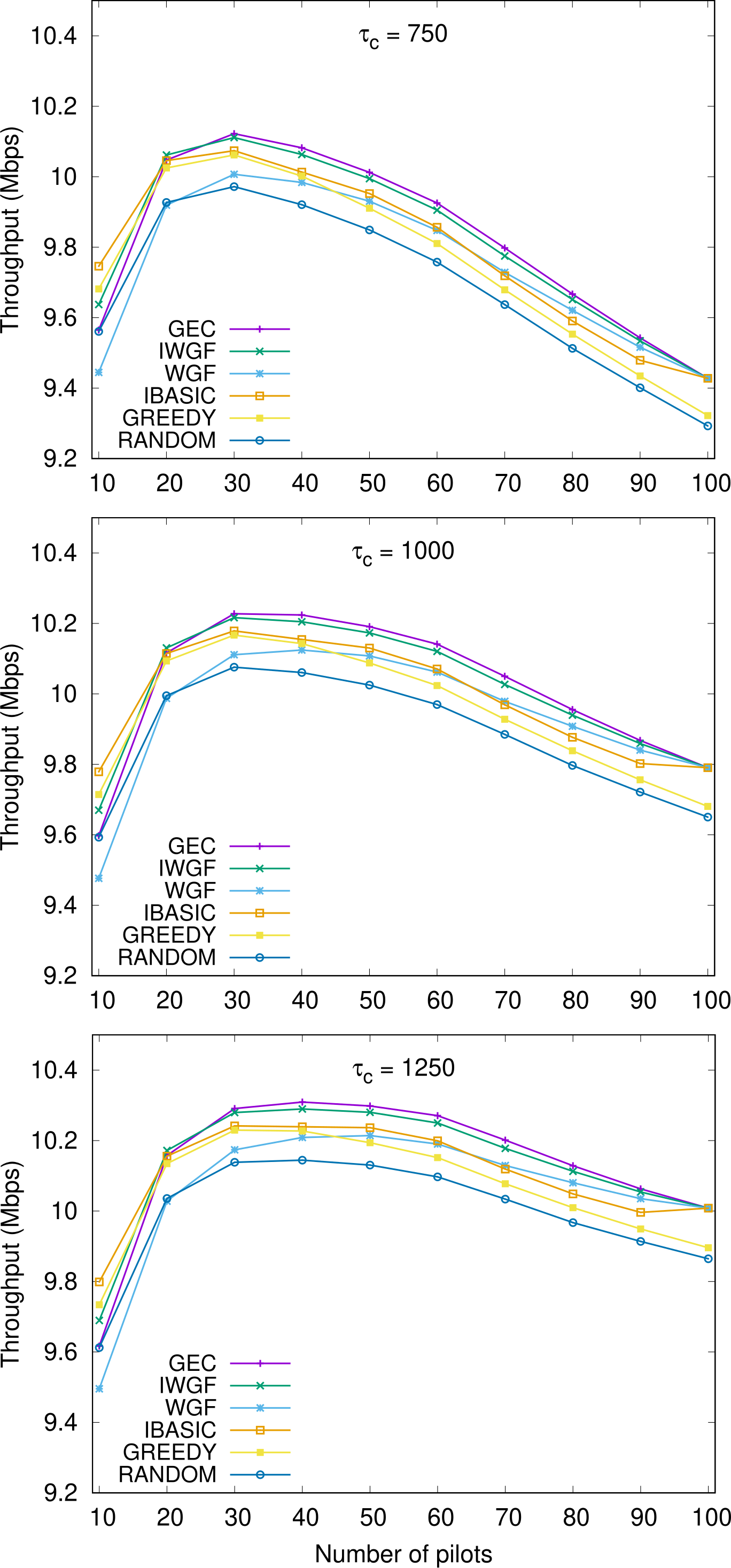}
\caption{Throughput $R^\mathrm{u}$ vs.\ number of pilots $P$.
Confidence-interval bounds at the $95\%$ level are about $\pm 0.3\%$ of the
average and occur for WGF at $P=10$. This percentage varies with
$\tau_\mathrm{c}$ in the order of $10^{-11}$.}
\label{figrate}
\end{figure}

For each value of $P\le K$, every result we report is an average over $10^4$
random trials, each beginning with the independent sampling of coordinates for
all $M$ APs and all $K$ users, and of values for all $z_{mk}$'s. The resulting
instance of the pilot-assignment problem is then submitted to GEC and five other
algorithms: an Improved WGF (IWGF) that uses the edge weights in
Eq.~(\ref{weight-init}), the original WGF, IBASIC with
$\delta=\max\{5,\lceil K/P\rceil\}$,\footnote{$\delta=5$, as in \cite{qzj22},
unless $5P<K$, in which case $\delta=\lceil K/P\rceil$.}
GREEDY,\footnote{The performance measure used by GREEDY (see
Section~\ref{contr}) is based on Eq.~(\ref{sinr}), so during pilot assignment
with GREEDY we use $\eta_k=1$ for every user $k$ \cite{naylm17}.} and RANDOM.
Our results are given in Figures~\ref{figsinr} and~\ref{figrate}, respectively
for $\mathrm{SINR^u}$ and $R^\mathrm{u}$ as functions of $P$. We omit confidence
intervals from the figures but inform their bounds in the figures' captions.

All plots suggest the superiority of GEC beginning at $P\approx 25$, followed
by IWGF, then variously by IBASIC, GREEDY, or WGF, though GREEDY is outperformed
by IBASIC and WGF beginning at $P\approx 45$. Excluding GREEDY and RANDOM, all
methods perform equally for $P=K$, indicating that they correctly avoid pilot
contamination altogether whenever possible. In the case of GEC, this is easily
seen by noting that the loop in line~\ref{while} of Algorithm~\ref{alggec} is
never entered if $P\ge K$. In conformity with Eq.~(\ref{rate}), throughput is
seen to increase with $\tau_\mathrm{c}$ for fixed $P$, but for fixed
$\tau_\mathrm{c}$ decreases after peaking as $P$ continues to grow. These trends
can also be seen as affecting the channel's spectral efficiency on the uplink,
which is given by $2B^{-1}R^\mathrm{u}=R^\mathrm{u}\times 10^{-7}$ (Mbps/Hz).

\section{Conclusion}
\label{concl}

We attribute the superiority of both GEC and IWGF to their formulation as a MAX
$P$-CUT problem with edge weights that, unlike those used by the original WGF,
reflect the fundamental quantity underlying the rise of pilot contamination when
a pilot is assigned to more than one user. This much is a consequence of our
discussion in Section~\ref{intr} regarding the ultimate centrality of
Eq.~(\ref{var}) in the choice of edge weights. One might also have expected IWGF
to perform better than GEC, given that the former's approximation ratio
surpasses the latter's. Our results in Section~\ref{rslt} show quite the
opposite and this should work as a reminder of what such ratios really mean.
They are lower bounds on how close to an optimal result the heuristic in
question can get, but in general only experimentation can clarify how those
lower bounds get surpassed in each case. For the experiments at hand, clearly
GEC was able to surpass its ratio's lower bound enough to perform better than
IWGF on average. Thus, given that the two algorithms have the same computational
complexity under massive parallelism, GEC is in the end the better choice.

\subsection*{Acknowledgments}

This work was supported in part by Conselho Nacional de Desenvolvimento
Científico e Tecnológico (CNPq), in part by Coordenação de Aperfeiçoamento de
Pessoal de Nível Superior (CAPES), and in part by Fundação Carlos Chagas Filho
de Amparo à Pesquisa do Estado do Rio de Janeiro (FAPERJ). It was also supported
by MCTIC/CGI.br/São Paulo Research Foundation (FAPESP) through projects Slicing
Future Internet Infrastructures (SFI2) – grant number 2018/23097-3, Smart 5G
Core And MUltiRAn Integration (SAMURAI) – grant number 2020/05127-2, and
Programmable Future Internet for Secure Software Architectures (PROFISSA) –
grant number 2021/08211-7.

\bibliography{cellfree-abbrev}
\bibliographystyle{unsrt}

\end{document}

%% file: table1.tex
\begin{tabular}{ccc}
\hline
$D=10^3$~m & $d_0=10$~m & $d_1=50$~m \\
$f=1.9\times 10^3$~MHz & $h_\mathrm{AP}=15$~m & $h_\mathrm{user}=1.65$~m \\
$\sigma_\mathrm{sf}=8$~dB & $\rho_\mathrm{p}=1.57\times 10^{11}$ & $\rho_\mathrm{u}=1.57\times 10^{11}$ \\
$B=2\times 10^7$~Hz & $\tau_\mathrm{c}=750,1000,1250$ & $P=\tau_\mathrm{p}\le 100$ \\
\hline
\end{tabular}